\documentclass[usenatbib]{mn2e}
\usepackage{graphicx}

\title[Mid-infrared imaging and modelling of IRAS 19500-1709]
{Mid-infrared imaging and modelling of the dust shell
around post-AGB star HD 187885 (IRAS 19500-1709)}

\author[K. L. Clube and T. M. Gledhill]{K. L. Clube$^{1}$\thanks{E-mail:
kclube@star.herts.ac.uk} and T. M. Gledhill$^{1}$\\
$^{1}$Dept. of Physics, Astronomy \& Maths, University of
Hertfordshire, College Lane, Hatfield, UK, AL10 9AB}
\begin{document}


\pagerange{\pageref{firstpage}--\pageref{lastpage}} \pubyear{2003}

\maketitle

\label{firstpage}

\begin{abstract}
We present 10 and 20~$\mu$m images of IRAS 19500-1709 taken with the
mid-infrared camera, OSCIR, mounted on the Gemini North Telescope. An
extended circumstellar envelope is detected, with the N band image
indicating an elongation in a NE-SW direction.  We use a dust
radiation transport code to fit the spectral energy distribution from
UV to sub-mm wavelengths, with a detached dust shell model.  A good fit
is achieved using dust composed of amorphous carbon, silicon carbide
and magnesium sulphide. We derive estimates for the inner and outer
radius, density and mass of the dust in the shell. The inner radius is
not resolved in our OSCIR imaging, giving an upper limit of 0.4
arcsec. With this constraint, we conclude that IRAS 19500-1709 must be
at least 4~kpc away in order to have the minimum luminosity
consistent with a post-AGB status.  
\end{abstract}


\section{Introduction}

Post-AGB stars are evolved stars with low to intermediate mass
(0.8-8.0~M$_{\odot}$) progenitors, that have left the Asymptotic Giant
Branch (AGB) but not yet become the central stars of Planetary Nebulae
(PN) (see review by van Winckel 2003). The heavy mass-loss that occurred on the AGB
has ended and the dust envelope that surrounds the star, as a result
of this mass-loss, expands away from the star, decreasing in
temperature while the temperature of the central star
increases. During this relatively short (10$^{3}$-10$^{4}$ years)
phase, these cool dust rich envelopes radiate most of their energy
at mid- and far-infrared wavelengths. Observations in the
mid-infrared probe the inner, warmest part of the envelope, the area
of most recent mass-loss which occurred near the end of the AGB.
Studies of individual objects in this transitional post-AGB phase
can be used to investigate this final period of AGB mass-loss and to
determine the physical and chemical properties of the dust.

IRAS 19500-1709 is associated with the high galactic latitude F2-3I
(Parthasarathy, Pottasch \& Wamsteker 1988) post-AGB star HD~187885
(SAO 163075) and has the double-peaked spectral energy distribution
(SED) typical of post-AGB stars with detached circumstellar dust
envelopes (Hrivnak, Kwok \& Volk 1989). The radiation from the star,
reddened by the dust shell, peaks at around 1~$\mu$m whereas the
thermal emission, from the shell itself, peaks at 25-30~$\mu$m. The post-3rd-dredge-up
 status of this object is confirmed by its high abundance of s-process elements (van
 Winckel \& Reyniers 2000).
 The
expansion velocity of the envelope, based on the CO-line emission, is
11~km~s$^{-1}$ with wings up to 30~km~s$^{-1}$ (Likkel et
al. 1987). The envelope is carbon rich, showing no OH or H$_{2}$O
maser emission (Likkel 1989). It has a weak 21~$\mu$m feature
(Justtanont et al. 1996) and a broad feature around 30~$\mu$m (Hony,
Waters \& Tielens 2002). The origin of these two features is
unknown. von Helden et al. (2000) identified titanium carbide (TiC) as
the carrier of the 21$\mu$m feature but this has recently been
disputed by Li (2003). Speck \& Hofmeister (2004) have suggested that
SiC may carry the 21~$\mu$m feature. Goebel \& Moseley (1985) proposed
that the 30~$\mu$m feature may be due to magnesium sulphide (MgS). The
central star has a temperature of 8000~K (van Winckel \& Reyniers
2000). Imaging polarimetry has shown that IRAS 19500-1709 has a
bipolar structure in scattered light (Gledhill et al. 2001).

In section 2 of this paper we present 10 and 20~$\mu$m images of IRAS
19500-1709 taken with the mid-infrared camera, OSCIR, on Gemini
North\footnote{Based on observations obtained at the Gemini
Observatory, which is operated by the Association of Universities for
Research in Astronomy, Inc; under a cooperative agreement with the
NSF on behalf of the Gemini partnership:the National Science
Foundation (United States), the Particle Physics and Astronomy
Research Council (United Kingdom), the National Research Council
(Canada), CONICYT (Chile), the Australian Research Council
(Australia), CNPq (Brazil) and CONICET (Argentina).}. In section 3 we
discuss the results of our modelling of these images and of the SED
using a dust radiation transport (RT) code.

\section[]{Observations}

We present 10 and 20~$\mu$m (N and Q3 band) images of IRAS 19500-1709
taken on 2001 July 12 with OSCIR, mounted on the 8.1-m Gemini North
Telescope (Fig.~1). OSCIR is a mid-infrared camera and spectrometer
system built by the University of Florida Infrared Astrophysics Group.
The array field of view is 11 x 11 arcsec and the pixel scale 0.089
arcsec. We used the standard nodding/chopping mid-infrared
observational technique to correct for the thermal background with a
chop frequency of 3 Hz and a chop throw of 15 arcsec. For each filter,
frame times were 10~ms and on-source exposure times were 4.3 min. For
flux calibration we used Vega, which we observed before and after
observing IRAS 19500-1709. Atmospheric transmission, as determined by
the Vega measurements, varied throughout the night and average
calibration factors have been calculated. The resultant N and Q3 band
fluxes, obtained by summing over the whole object, are given in
Table~1. We have also used the Vega observations as an indication of
the point spread function (PSF) and the full width at half-maximum
(FWHM) is shown in Table 1 along with other observing details.

Flat-field images were obtained for each filter by exposing on sky
and on a polystyrene flat-field source without chopping and nodding
and the flat-field constructed using the {\sevensize OFLAT} task,
which is part of the {\sevensize GEMINI} package within {\sevensize
IRAF}\footnote{Image Reduction and Analysis Facility is distributed by
the National Optical Astronomy Observatories, which is operated by the
Association of Universities for Research in Astronomy, Inc., under
cooperative agreement with the National Science Foundation}. The
observations were reduced using the {\sevensize OREDUCE} task in the
{\sevensize GEMINI IRAF} package which derives the average of the chop
and nod differences and produces a single flat-field-corrected image
for each filter. A few lines of bad pixels on these images, caused by
an OSCIR channel problem which was present during the early Gemini
observing runs, were fixed using an IDL procedure provided by Scott
Fisher (OSCIR support team).

The images of IRAS 19500-1709 show that it is extended relative to the
PSF, with a FWHM of 1.2 arcsec in both filters and diameter of 4
arcsec.  The N band image appears elongated in the outer contours
with position angle (PA) of
$20\pm10$ deg. We note that the telescope PSF, as determined by
observations of Vega, is also elongated in a similar direction (shown
contoured in Fig.~1). However, the outer contours of the Q3 band image
appear round, although the PSF is also elongated, suggesting that the
elongation seen in the outer regions of the N band image may be
real. This interpretation is strengthened by previous lower spatial
resolution imaging which also tentatively suggests an elliptical
structure with PA $30\pm10$ deg. (Meixner et al. 1997).  The
brightness peak in the N band image appears approximately 0.2 arcsec
off centre to the east, relative to the outer contours. The inner
boundary, of any detached shell, is unresolved.

\begin{table}
\caption{Measurements for IRAS 19500-1709 for each filter showing the
central wavelength ($\lambda_c$), bandpass ($\Delta\lambda$), flux and
the FWHM of the PSF, estimated from observations of Vega. Due to
variations in sky transparency, the errors on the fluxes may be as
large as $\pm50$ per cent.}
\begin{tabular}{@{}lcccc@{}}
\hline
Filter & $\lambda_c$ & $\Delta\lambda$ & Flux & FWHM (Vega) \\
  & ($\mu$m) & ($\mu$m) & (Jy) & (arcsec) \\
\hline
N wide & 10.75 & 5.23 &  33 & 0.51 \\
Q3 & 20.80 & 1.65 & 134 & 0.64 \\
\hline 
\end{tabular}
\end {table}


\begin{figure*}
\includegraphics[width=140mm]{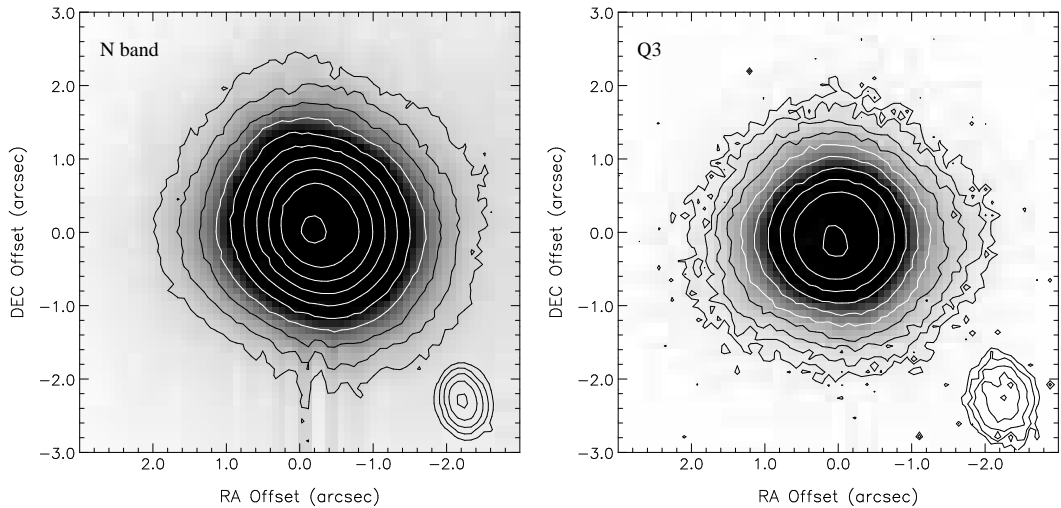}
\caption{Images of IRAS 19500-1709 from the OSCIR camera on Gemini
North using N and Q3 filters. The PSF is shown contoured in the lower
right corners. All innermost contours are 95\% of the peak values and
spaced at 0.5 magnitude increments.} 
\end{figure*}

\section{Modelling and Discussion}

We use the dust RT code originally developed by Efstathiou \&
Rowan-Robinson (1990), which solves the radiation transfer equation
for an axisymmetric distribution of dust around a central source. The
code allows up to seven different grain materials and has been
modified to include a power-law size distribution of dust grains
(Gledhill \& Yates 2003). Although the code is fully 2-D, we use a
spherically symmetric dust shell to model 19500-1709 since there is
only tentative evidence for axisymmetry in the OSCIR observations and
this would not provide sufficient constraints to justify the inclusion
of the additional parameters required by a 2-D model.

The input parameters to a spherical shell model include the ratio of
the stellar radius to the inner shell radius ($r_{*}/r_{1}$), the
ratio of the inner shell radius to the outer shell radius
($r_{1}/r_{2}$), and the stellar temperature ($T_{\rm eff}$). The dust
density is specified by the extinction ($A_{\rm v}$) through the shell
with the dust density assumed to decline with radius as $r^{-2}$,
corresponding to a constant mass-loss rate. The absorption and
scattering efficiencies are calculated for each grain type using Mie
theory with the dust properties determined by the optical constants
and grain sizes. We assume spherical dust grains with radii between
$a_{\rm min}$ and $a_{\rm max}$ with a size distribution of $n(a)
\propto a^{-q}$.

The resultant model spectrum is plotted against the observational data
for IRAS 19500-1709, which have been taken from the literature. These
include optical and near-infrared photometry (Hrivnak et al. 1989), a
7.6-23.6~$\mu$m UKIRT spectrum (CGS3) provided by K. Justtanont, {\it
IRAS} fluxes, 8.5, 10 and 12.2~$\mu$m flux (Meixner et al. 1997), {\it
IUE} data\footnote{Based on INES data from the {\it IUE} satellite},
{\it ISO} SWS\footnote{{\it ISO} SWS06 observing program
rszczerba-PPN30} and LWS\footnote{{\it ISO} LWS01 observing program
mbarlow-dust 3} spectra and JCMT flux (van der Veen et al. 1994).

\subsection{Fitting the SED}

Around 90 models were run and the best-fitting model, as assessed by
eye, is shown in Fig. 2. The final model includes dust composed of
amorphous carbon (amC), silicon carbide (SiC) and magnesium sulphide
(MgS) grains. The amC dust fits the continuum of the SED quite well.

The broad emission feature, extending from 10-13~$\mu$m and peaking
around 12~$\mu$m was observed by Justtanont et al. (1996) but has not
been conclusively identified.  Although SiC is thought to be a common
dust component in carbon stars (Speck et al. 1997), it results in a
feature which peaks at 11.3~$\mu$m and is too narrow to fit the broad
12~$\mu$m `plateau' feature.  However, in the absence of an
alternative identification, we have used $\alpha$-SiC optical
constants (P\`{e}gouri\`{e} 1988) to produce a feature in this
wavelength range. The 8-10~$\mu$m region is also problematic, in that
it exhibits unidentified features. Kwok, Volk \& Bernath (2001) suggest
 that these 8 and 12~$\mu$m plateau features may be due to aromatic and
 aliphatic hydrocarbons. Justtanont et al. (1996) propose that
 a polycyclic aromatic hydrocarbon (PAH), such as chrysene
(C$_{18}$H$_{12}$), provides the best match to UIR features in the
8-13~$\mu$m region. We have briefly explored PAH materials using the
cross-sections of Li \& Draine (2001) but find that the features
produced are too narrow to match the plateau-nature of the observed
emission and additional features in the 16-25~$\mu$m region of the
spectrum are produced, which are not seen in the data.

We include MgS as being the possible carrier of the 30~$\mu$m band. It
has been found to provide a reasonable fit to this feature (Hony et
al. 2002). We used optical constants for Mg$_{x}$Fe$_{(1-x)}$S from
Begemann et al (1994) with $x$=0.9, as an approximation to MgS (also
see Szczerba et al. 1997, Hony et al. 2003). These data do not extend
below 10~$\mu$m however, so it was necessary to approximate the optical
and UV extinction cross-sections to ensure that the MgS grains absorb
sufficient radiation at short wavelengths to radiate in the
mid-IR. This was done by adopting the optical constants of parallel
graphite (Draine \& Lee 1984) below 4~$\mu$m and interpolating between
4 and 10~$\mu$m to produce a smooth extinction curve. The model shows a
 feature in the UV part of the spectrum, between 0.2-0.25~$\mu$m. Both parallel graphite
 and amC show structure in their extinction cross sections in this region.
 The feature around 30~$\mu$m is broader and peaks at a longer wavelength than that
produced by our model. This may be because we have used spherical
grains. Variations in grain shape can influence the emission profile,
with ellipsoidal grains producing a broader peak (Hony et al. 2002).
Szczerba et al. (1999) suggest that the 30~$\mu$m feature may be
related to the 21~$\mu$m feature, as stars with the 21~$\mu$m feature
seem to have the 30~$\mu$m feature as well. The reverse is not the case
though; the 21~$\mu$m feature is very weak in 19500-1709, which van
Winckel \& Reyniers (2000) suggest may be due to the temperature of this
 object as the feature is observed to be stronger in the cooler 21~$\mu$m
 post-AGB stars in their sample. It certainly seems
possible that the 30~$\mu$m feature may be due to an unknown
carbonaceous component, rather than MgS, as it is a feature which is
only seen in carbon rich objects.
 
In our model the central star is assumed to radiate as a black body with
$T_{\rm eff}=8000$~K. This temperature is higher than has been 
previously assumed in some of the
literature (e.g. Hrivnak et al. (1989), G\"{u}rtler, K\"{o}mpe \&
Henning (1996), Meixner et al. (1997) use temperatures of 7000-7500~K 
which have been estimated from the spectral type). However
detailed analysis of spectral line data suggests a higher
temperature of 8000~K (van Winckel \& Reyniers (2000)) so we adopt
this as the most reliable estimate. Parthasarathy
et al. (1988) suggested a temperature of 8500~K might be
required to fit the UV spectrum of IRAS 19500-1709 which they found
to be peculiar because of a broad absorption feature around
0.17~$\mu$m. 
Increasing the temperature of the central star improves the fit around
the UV part of the SED but also increases the prominence of the SiC
emission feature. So an increase in temperature needs to be balanced
with a decrease in the abundance of SiC. However if the abundance of
amC is too high the model does not fit the data around 15-25~$\mu$m. A
dust mixture comprising 40 per cent amC, 30 per cent SiC and 30 per
cent MgS provided the best fit to the data.

A value of $A_{\rm v}$=1.2 provided the best fit around the UV and
optical part of the spectrum. Increasing $A_{\rm v}$ results in too
much extinction in this region and also too much mid-IR flux. Larger
dust grains produce too much flux in the UV and optical region and too
strong an emission feature around 11~$\mu$m but insufficient flux in
the mid-IR peak. A value of $a_{\rm min}$=0.01~$\mu$m yielded the best
results along with a grain size power-law index of $q$=5. Reducing the
value of $q$ includes too many larger grains and so also results in a
mid-IR peak which is too weak. 

The fit to the 10-25~$\mu$m region is strongly affected by
$r_{*}/r_{1}$ which determines the temperature of the warmest dust. A
value of $8\times10^{-5}$ provided the best fit. Increasing this value
results in dust grains which are too hot and a poor fit around this
region. Reducing the value of $r_{*}/r_{1}$ results in dust which is
too cold and so not enough flux at mid-IR wavelengths and too much in
the 40-160~$\mu$m region. The ratio of the inner shell radius to the
outer shell radius ($r_{1}/r_{2}$) is determined to be 0.1 as this
produces sufficient far-IR emission to fit the data longward of
100~$\mu$m. Extending the boundary any further than this produces too
much far-IR emission.

After fitting the model to the data, the model curve was scaled to
fit the observed flux. Table 2 shows the parameters for the model fit
along with the derived parameters.

\begin{figure}
\includegraphics[width=90mm]{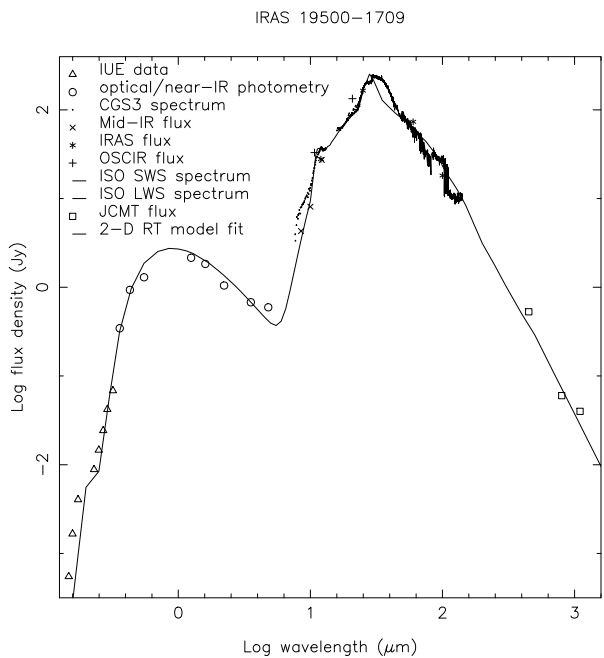}
\caption{SED and model fit (solid black line).}
\end{figure}
  
\subsection{Derived parameters and the distance to 19500-1709}

The profiles of the raw model images are double-peaked showing a
resolved inner boundary. The inner boundary of the OSCIR images,
however, is unresolved. In order to match the model to the data and so
determine the scale of the model (i.e. the size of a model pixel in
arcsec), the model images were smoothed using a Gaussian filter until
they resembled the OSCIR images. Using smoothing steps of 1 model
pixel, the Gaussian smoothing PSF was determined to have FWHM $>$ 11
pixels. A value less than this resulted in a dip in the profile so
that FWHM=11 is the lowest value to show a single peak in the
profile. Equating this degree of smoothing to the observed PSF (see Table~1) gives a model pixel size of $0.046$ arcsec. This
then leads to values of $r_{1}<0.4$ arcsec and $r_{2}<4$ arcsec (for
$r_{1}/r_{2}=0.1$). Since $r_{*}/r_{1}=8\times10^{-5}$ then $r_{*}<6.87$~R$_{\odot}$ and $L<173$~L$_{\odot}$. These and other resulting upper limits on the derived parameters
are listed in Table~2 for a distance of 1~kpc.



Sch\"{o}nberner (1983) determined lower limits for the mass and
luminosity of central stars of PN of approximately 0.55~M$_{\odot}$ and
2500~L$_{\odot}$. If IRAS 19500-1709 is to be this luminous then it must be at a distance of at least 4~kpc.
 Pottasch \&
Parthasarathy (1988) conluded that $D=5.5$~kpc based on the optical-IR
energy balance.  The value of $r_{*}/r_{1}$ is well constrained by the
model fit: increasing or decreasing this value results in dust that is
either too hot or too cool, respectively. We explored the possiblility
that axisymmetric models might allow us to vary $r_{*}/r_{1}$, by
redistributing the dust into a smaller volume of the shell, however
the effect was minimal. Meixner et al. (1997) have used a larger value
of $r_{*}/r_{1}=2.7\times10^{-4}$ but it is evident from their SED fit
that the dust is too hot. With the model constraining $r_{*}/r_{1}$ in
this way we conclude that IRAS 19500-1709 must be at least 4 kpc away for it to have the minimum 
luminosity consistent with a post-AGB status and
this distance is adopted in the rest of the paper.

Dividing $r_{1}$ by the expansion velocity of the envelope ($11$~km~s$^{-1}$,
Likkel et al. 1987), shows that, for $D=4$~kpc, the shell became detached
(i.e. the high mass loss phase ended) at least 690 years ago. 
This is comparable to the result of 740 years obtained by Meixner et al. 
(1997). The age of the shell, calculated by dividing $r_{2}$ by the expansion velocity, is at least 6900 years.

The mass of dust in the shell and the grain number density at $r_{1}$
were calculated using an average of the densities of the 3 grain types
used in the model and an average value for the extinction cross
sections. Using the dust shell dimensions derived from the model fit,
then to produce an extinction through the dust shell of $A_{\rm
v}=1.2$, which is needed to fit the SED, requires a dust number
density at the inner shell radius of $n_{1}=5.6\times10^{-5}$~cm$^{-3}$,
assuming $D=4$~kpc.  Using this number density, assuming a $r^{-2}$
dust density distribution and an average grain bulk density of 
2.56~g~cm$^{-3}$ results in a total mass of dust of $M_{\rm
d}=1.89\times10^{-3}$~M$_{\odot}$. Gledhill, Bains \& Yates (2002) used 
their 850~$\mu$m SCUBA photometry to obtain a lower limit of
$M_{d}>1.6\times10^{-3}$~M$_{\odot}$ at 4~kpc.

Using a dust-to-gas mass ratio of $4.5\times10^{-3}$, an approximate
value used for carbon-rich AGB stars (Jura 1986), the total envelope
mass at $D=4$~kpc is $0.42$~M$_{\odot}$. According to our calculations
the mass loss episode, responsible for the shell, lasted 6200 years so
this results in an average mass loss rate of
$6.8\times10^{-5}$~M$_{\odot}$~yr$^{-1}$ at this distance. This is
comparable to the mass loss rate ($\dot{M}$) derived by G\"{u}rtler et
al. (1996) of $6\times10^{-5}$~M$_{\odot}$~yr$^{-1}$ using a RT code
and amC dust grains. It lies between the values derived by Meixner et
al. (1997, $\dot{M}=1.4\times10^{-4}$~M$_{\odot}$~yr$^{-1}$) and
Hrivnak et al. (1989,
$\dot{M}=1.5\times10^{-5}$~M$_{\odot}$~yr$^{-1}$). Omont et al. (1993)
calculate a value of
$\dot{M}=1.2-1.5\times10^{-5}$~M$_{\odot}$~yr$^{-1}$ from CO line
emissions. All of these rates are for $D=4$~kpc.


\subsection{Evidence for axisymmetry}

Near-IR imaging polarimetry observations by Gledhill et al. (2001) show
evidence for two scattering peaks on either side of the star, which suggest
scattering from the inner edge of an axisymmetric envelope. The line joining
the peaks is at PA$\sim100\degr$, approximately orthogonal to the elongation
in the N band image of Fig.~1. The morphology appears remarkably similar to 
that described in mid-IR and near-IR imaging of 17436+5003 (Gledhill \& Yates 
2003; Gledhill et al. 2001). Although there is insufficient information in our
OSCIR images to justify axisymmetric RT models, we suggest that 19500-1709
has an optically thin detached axisymmetric dust shell, resulting from enhanced
equatorial mass-loss. 




\begin{table}
\caption{Model and derived parameters resulting from the final fit 
to the SED.}
\begin{tabular}{@{}cc@{}}
\hline
Parameter & Value \\
\hline
$r_{*}/r_{1}$ & 8x10$^{-5}$ \\
$r_{1}/r_{2}$ & 0.1 \\
$A_{\rm v}$ (mag) & 1.2 \\
$q$ & 5 \\
$a_{\rm min}$ ($\mu$m) & 0.01 \\
$a_{\rm max}$ ($\mu$m) & 2 \\
$T_{\rm eff}$ (K) & 8000 \\
Grain type \\
amC (\%) & 40 \\
SiC (\%) & 30 \\
MgS (\%) & 30 \\
Grain temp. (min) \\
amC (K) & 53 \\
SiC (K) & 48 \\
MgS (K) & 42 \\
Grain temp. (max) \\
amC (K) & 170 \\
SiC (K) & 155 \\
MgS (K) & 125 \\
{\em Derived parameters} \\
$r_{1}$ (cm/kpc) & $<5.98$x10$^{15}$ \\
$r_{2}$ (cm/kpc) & $<5.98$x10$^{16}$ \\
$r_{*}$ (R$_{\odot}$/kpc) & $<6.87$ \\
$L$ (L$_{\odot}$/kpc$^{2}$) & $<173$ \\
$n_{1}$ (cm$^{-3}$/kpc$^{-1}$) & $>2.2$x10$^{-4}$ \\
$M_{\rm d}$ (M$_{\odot}$/kpc$^{2}$) & $<1.16$x10$^{-4}$ \\
\hline
\end{tabular}
\end{table}

\section{Conclusions}

We present mid-infrared images and a model of the SED of IRAS
19500-1709 using amC, SiC and MgS grains. We find that small dust
grains ($a_{\rm min}=0.01$~$\mu$m) and a steep power law size distribution
($q$=5) provide the best fit to the data. The observations constrain
the inner radius of any detached dust shell to be a maximum size of 0.4 arcsec. Our results suggest that this object must be at a
distance of at least 4 kpc to be sufficiently luminous to be a
post-AGB star. We show that, at this distance, IRAS 19500-1709 would
have an envelope mass of 0.42~M$_{\odot}$, that during the high
mass-loss phase the mass-loss rate was
$6.8\times10^{-5}$~M$_{\odot}$~yr$^{-1}$ and that this phase ended
$\sim700$~yrs ago. Although the inner envelope structure has not been
resolved in these OSCIR observations, there is evidence for
axisymmetry along an axis perpendicular to that seen in near-IR
imaging observations.

\section*{Acknowledgments}

This research has made use of the SIMBAD database, operated at CDS,
Strasbourg, France. All model calculations were run on the HiPerSPACE
Computing Facility at University College, London. We thank
K. Justtanont for providing the UKIRT CGS3 spectrum. Scott Fisher
 is thanked for his help with the data. K. Clube
is supported by a PPARC studentship.

\bsp

\label{lastpage}

\end{document}